\documentclass[journal=jacsat,manuscript=letter, 
]{achemso}

\usepackage[version=3]{mhchem} 
\usepackage[T1]{fontenc}
\usepackage{hyperref}
\usepackage{color}
\usepackage{bm}
\usepackage{bbold}
\usepackage{amsmath}
\usepackage{amssymb}
\usepackage{mathptmx}
\usepackage{mathrsfs}
\usepackage[normalem]{ulem}

\usepackage{siunitx}

\newcommand{\Angstrom}{\mbox{\normalfont\AA}}

\author{Connor L. Box}
\affiliation{Department of Chemistry, University of Warwick, Gibbet Hill Road, CV4 7AL, Coventry, UK.}
\author{Nils Hertl}
\affiliation{Department of Chemistry, University of Warwick, Gibbet Hill Road, CV4 7AL, Coventry, UK.}
\author{Wojciech G. Stark}
\affiliation{Department of Chemistry, University of Warwick, Gibbet Hill Road, CV4 7AL, Coventry, UK.}
\author{Reinhard J. Maurer}
\affiliation{Department of Chemistry, University of Warwick, Gibbet Hill Road, CV4 7AL, Coventry, UK.}
\alsoaffiliation[Second University]
{Department of Physics,  University of Warwick, Gibbet Hill Road, CV4 7AL, Coventry, UK.}
\email{r.maurer@warwick.ac.uk}

\title{Room Temperature Hydrogen Atom Scattering Experiments Are Not a Sufficient Benchmark to Validate Electronic Friction Theory}

\begin{tocentry}
  \includegraphics[width=3.3in]{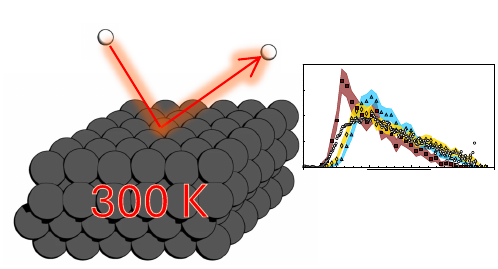}
\end{tocentry}
\begin{document}

\begin{abstract}
In the dynamics of atoms and molecules at metal surfaces, electron-hole pair excitations can play a crucial role. In the case of hyperthermal hydrogen atom scattering, they lead to nonadiabatic energy loss and highly inelastic scattering. Molecular dynamics with electronic friction simulation results, based on an isotropic homogeneous electron gas approximation, have previously aligned well with measured kinetic energy loss distributions, indicating that this level of theoretical description is sufficient to describe nonadiabatic effects during scattering. In this study, we demonstrate that friction derived from density functional theory linear response calculations can also describe the experimental energy loss distributions, although agreement is slightly worse than for the simpler isotropic homogeneous electron gas approximation. We show that the apparent agreement of the homogeneous electron gas approximation with experiment arises from a fortuitous cancellation of errors as friction is overestimated close to the surface and the spin transition is neglected. Differences in frictional treatment affect single, double, and multi-bounce scattering trajectories in distinct ways, altering the shape of low-temperature energy loss distributions. These distinctions are largely absent at room temperature, but may be measurable in future low-temperature scattering experiments may be able to measure.
\end{abstract}


The scattering of hyperthermal hydrogen atoms with metal surfaces is a well-studied phenomenon, known to cause substantial energy transfer to the surface~\cite{bunermann_electron-hole_2015}. At room temperature, the energy loss distributions (ELDs) of scattering hydrogen atoms are broad, lacking distinct features with only subtle differences between metal surfaces~\cite{dorenkamp_hydrogen_2018, hertl_electronically_2022}. Classical molecular dynamics (MD) simulations underestimate this energy transfer because they do not account for nonadiabatic effects during the dynamics and the excitation of electron-hole pairs (EHPs) in the metal.\cite{bunermann_electron-hole_2015, janke_accurate_2015} To address this shortcoming, several simulation methods have been developed, among which molecular dynamics with electronic friction (MDEF) stands out as a practically feasible approach. MDEF has been applied to the dynamics of atoms~\cite{janke_accurate_2015,askerka_role_2016} and molecules~\cite{maurer_mode_2017,galparsoro_hydrogen_2016,spiering_testing_2018, spiering_orbital-dependent_2019, zhang_hot-electron_2019, box_determining_2021} at metal surfaces, including under the influence of laser pulses.~\cite{fuchsel_dissipative_2011,scholz_femtosecond-laser_2016} Theoretical studies applying MDEF simulations to H atom scattering~\cite{Janke_PhD_2016,kandratsenka_unified_2018,martin_h_2022} were able to primarily link scattering events with low energy loss to single bounce scattering of hydrogen atoms, while events that exhibit high energy losses are associated with atoms that experience multiple bounces.

MDEF simulates the dynamics on the adiabatic ground state potential energy surface while introducing a drag force on the nuclear motion, that accounts for the energy transfer to EHP excitations.\cite{headgordon_molecular_1995} Remarkably, MDEF simulations have successfully reproduced the ELDs for hydrogen atom scattering from a range of metal surfaces.\cite{janke_accurate_2015,dorenkamp_hydrogen_2018, hertl_electronically_2022} This success, however, is puzzling for several reasons. 

Firstly, in the context of hydrogen atom scattering from a metal surface, the unpaired spin of the hydrogen atom is screened in close vicinity of the metal surface. Upon approaching the surface, the occupied and unoccupied states of the hydrogen atom hybridise with the metal electrons leading to an increase in their spectral width. The spin states become degenerate, and the spin-polarisation is lost. This gives rise to a strong nonadiabatic effect that is typically beyond the scope of MDEF.\cite{mizielinski_electronic_2005} Indeed, it was previously shown that electronic friction evaluated within time-dependent perturbation theory in the Markov approximation,  also referred to as orbital-dependent friction (ODF)~\cite{luntz_comment_2009,askerka_role_2016,spiering_testing_2018}, gives rise to a sudden rise in stopping power at the spin transition geometry.\cite{ mizielinski_electronic_2005, trail_electronhole_2003} While this effect is "broadened out" in nonadiabatic mean-field (Ehrenfest) dynamics due to memory  effects,~\cite{lindenblatt_ab_2006} its contribution to the energy loss is similar and significant in both cases.~\cite{mizielinski_newnsanderson_2008} Previous MDEF studies of hyperthermal hydrogen scattering based on the local density friction approximation (LDFA) \cite{janke_accurate_2015,dorenkamp_hydrogen_2018, hertl_electronically_2022} have effectively ignored this spin transition. LDFA  evaluates electronic friction within a homogeneous electron gas approximation and neglects the electronic structure and spin state of the adsorbate. As a result, previous LDFA  studies are expected to underestimate the energy loss in the region of the spin transition, whereas ODF calculations will describe the spin transition in the extreme quasi-adiabatic limit. Both do not fully capture how this effect depends on the details of the trajectory.

Secondly, while the magnitude of electronic friction from LDFA and ODF align for the homogeneous electron gas~\cite{hellsing_electronic_1984}, the validity of LDFA based on DFT electron densities of clean metal surfaces has been previously questioned\cite{luntz_comment_2009}.

LDFA data presented in studies of hydrogen atoms and molecules chemisorbed at metal surfaces ~\cite{luntz_comment_2009,askerka_role_2016,spiering_testing_2018} overestimate electronic friction and underestimate vibrational lifetimes when compared to ODF. Notably, practical calculations of electronic friction based on ODF within DFT also provide challenges as they require explicit Fermi surface integration that is notoriously difficult to converge. Specific choices regarding mathematical expressions and numerical approximations can affect the results and have previously been discussed in great detail, motivating careful validation against experimentally accessible data.~\cite{box_ab_2023} The hydrogen atom vibrational lifetime on the surface can be measured experimentally and related to the electronic friction coefficient along the corresponding vibrational mode. Reported hydrogen relaxation times 
by LDFA for Pd(100) (0.2 ps)\cite{blanco-rey_electronic_2014}, Ru(0001) (0.2 ps)\cite{fuchsel_dissipative_2011}, Pb(111) (0.3 ps)\cite{saalfrank_vibrational_2014}, and Pt(111) (0.2 ps, \textit{vide infra}) are a factor of 3-4 faster than experimental measurements for hydrogen atom relaxation on Pt(111) (0.8$\pm$0.1 ps)\cite{palecek_characterization_2018} and Cu(111) ( 0.7 ps)\cite{lamont_dynamics_1995}. As the value of LDFA friction for chemisorbed hydrogen varies only little across close-packed surfaces of transition metals, this may point to an overestimation of friction for chemisorbed hydrogen on transition metals by LDFA. This will likely affect the predicted ELDs of scattered H atoms.

In this work, we simulate hyperthermal hydrogen scattering on Pt(111) for which experiments were previously reported.~\cite{dorenkamp_hydrogen_2018,lecroart_adsorbate_2021} We perform MDEF simulations with two electronic friction tensor models based on LDFA and full DFT linear response calculations, i.e., ODF.~\cite{maurer_ab_2016,box_ab_2023}  The two models exhibit very different configuration-dependence of friction, yet both predict room temperature ELDs that broadly agree with the experiment. DFT-based electronic friction predicts lower energy loss close to the surface than LDFA, which is in better agreement with time-resolved spectroscopy measurements, but only agrees with the experimental ELD if the spin transition is included. Simulations of low-temperature ELDs, however, reveal clear differences that arise from the specific features of the two friction profiles.


In MDEF, the dynamics are governed by the Langevin equation\cite{headgordon_molecular_1995} 
\begin{equation}
M \ddot{R}_{a}=-\frac{\partial V(\boldsymbol{R})}{\partial R_{ a}}-\sum_{a^{\prime}} \Lambda_{aa^{\prime}} \dot{R}_{ a^{\prime}}+\mathcal{R}_{a}(t)
\label{eq:GLE}
\end{equation}
where $V(\boldsymbol{R})$ represents the potential energy surface (PES) which is a function of all atomic positions, $\boldsymbol{R}$. We calculate elements of the electronic friction tensor, $\bm{\Lambda}$, for the three Cartesian coordinates, $a$, of the hydrogen atom projectile. $\mathcal{R}(t)$ is the random force, which establishes detailed balance at a given electronic temperature $T$ via the second fluctuation-dissipation theorem.\cite{kubo} The energy transfer between the translational degrees of freedom of the H atom and the electrons is governed by the second and third terms on the right hand side of Equation \ref{eq:GLE}. Within MDEF, the nuclei are treated as classical particles. ELDs from classical simulations of H atom scattering from Xe(111) are in excellent agreement with experiments at $\sim$40\,K which suggests that nuclear quantum effects do not strongly affect ELDs.~\cite{hertl_multibounce_2021}


We employ a previously reported~\cite{kammler_genetic_2017} high-dimensional effective medium theory\cite{emt1987,emt1996,emt2022} (EMT) PES for hydrogen scattering on Pt(111) that was parametrized with PBE-DFT data.~\cite{pbe1, pbe2} 
The EMT-PES reports a root mean squared error with respect to the DFT data of 259\,meV, yet good agreement between the ELDs obtained with this PES and the experiment has been reported.\cite{dorenkamp_hydrogen_2018} 
The shape of the energy loss distributions at room temperature has been shown to be insensitive to the nature and the shape of the underlying PES for H/W(110).\cite{hertl_random_2021,martin_h_2022}  We employ a (2$\times$2) surface slab model with six metal layers. Electronic friction in the LDFA is calculated from electron scattering phase shifts that arise for a hydrogen ion moving through a homogeneous electron gas with a density equivalent to the density at the position of the hydrogen atom in our simulations.~\cite{gerrits_electronic_2020} The resulting friction is a scalar coefficient per atom and only depends on the local electron density, which we evaluate based on the EMT-PES model density. \cite{janke_accurate_2015, kammler_genetic_2017} 

Gaussian process regression~\cite{rasmussen_gaussian_2005} models of ODF friction tensors were generated using \texttt{scikit-learn}~\cite{pedregosa_scikit-learn_2011} (version 1.2.0) with a vector of the inverse hydrogen atom distances to the platinum atoms as a descriptor. ODF tensors without spin polarization were calculated with the electronic structure code FHI-aims \cite{blum_ab_2009, box_ab_2023} (version 210928) for $3198$ configurations. The $757$ structures with the hydrogen atom above $1.8$~$\Angstrom$ from the surface were recalculated with system spin constrained to $^2\mathrm{S}_{1/2}$. The MDEF simulations were performed using the NQCDynamics.jl package (development version based on version 0.12) \cite{gardner_nqcdynamics_2022}. Further method details are provided in the Supporting Information (SI).



\begin{figure}[]
    \includegraphics[width=3.3in]{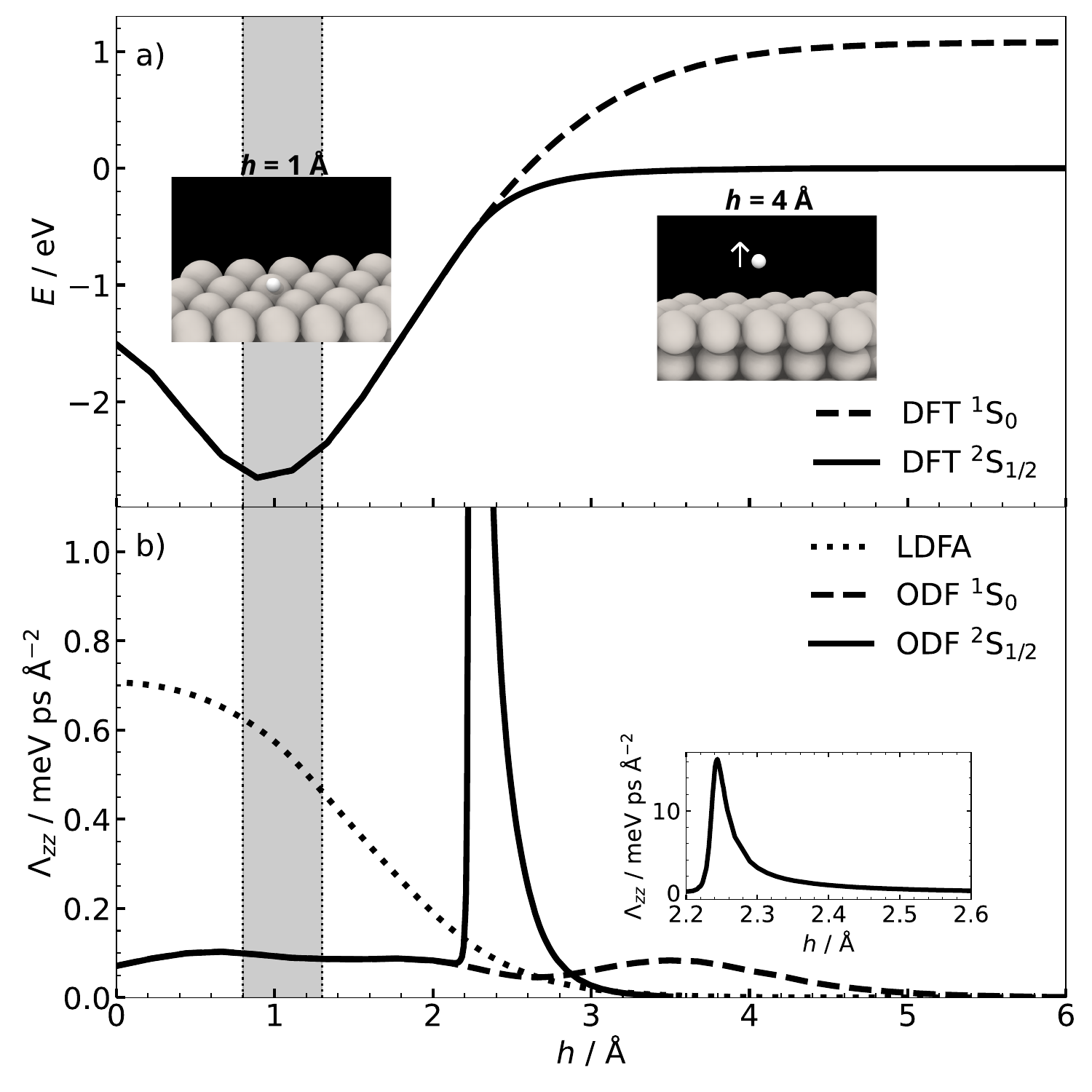}
    \caption{(a) The interaction energy of H on Pt(111) as a function of its atom-surface distance, $h$, above the fcc-hollow site. The solid and dashed lines represent the energy curves for the doublet-constrained spin-polarised and spin-unpolarised DFT calculations, respectively. (b) The friction models as a function of atom-surface distance. The dotted line represents the LDFA values obtained with density values. The dashed and solid lines represent the $\Lambda_{zz}$ values for the constrained spin-polarised and spin-unpolarised DFT calculations, respectively.  
    The inset of panel b) shows the entire spin curve for the constrained doublet state with different axis limits. The grey shaded areas in a) and b) indicate the distances relevant for vibrationally excited, chemisorbed H atoms.
    }
    \label{fig:Friction_map}
\end{figure}

The EMT-PES has been shown to provide a good fit to PBE-DFT data~\cite{lecroart_adsorbate_2021}, which for a hydrogen atom on Pt(111), predicts a binding well with an adsorption minimum in the fcc-hollow site at $0.87$~\AA{} (Figure~\ref{fig:Friction_map}a). At large atom-surface separation, the doublet state ($^2$S$_{1/2}$)  is more stable and the atom exhibits an unpaired spin. At closer distances, the $^2$S$_{1/2}$ and $^1$S$_0$ states become degenerate in energy as the spin is screened by the metal surface. The adsorption height, $h$, at which this occurs is approximately $2.3$~{\AA}, similar to previous findings for H/Cu(111)\cite{trail_electronhole_2003}. 

The three friction models differ in their position dependence as a function of atom-surface separation (Figure~\ref{fig:Friction_map}b). As LDFA friction only depends on the electron density of the clean metal substrate, it increases monotonically with decreasing atom-surface separation from a height of about 3~\AA{}. It reaches a value of around $0.5\,\mathrm{meV\,ps}$\,{\AA}$^{-2}$ at around 1~\AA{}. LDFA does not account for changes in the magnetisation density of the projectile~\cite{li_molecular-dynamics_1992}. The coefficient of the ODF friction tensor perpendicular to the surface ($\Lambda_{zz}$) is sensitive to the spin state of the system as the atom approaches the surface. In the $^2$S$_{1/2}$ state, ODF friction starts to rise at the same atom-surface distance as LDFA, but exhibits a sharp rise to approximately $15.6$~$\mathrm{meV\,ps}$\,{\AA}$^{-2}$ at the separation distance of 2.3\,{\AA} above the fcc-hollow site, where the spin is screened by the metal and doublet and singlet states become degenerate. Below this height, the friction values are identical for both spin states. The ODF $^1$S$_0$ extends much further into the vacuum, away from the surface, which is an artefact that arises from enforcing an incorrect spin state at these geometries.

The grey shaded region in Figure~\ref{fig:Friction_map} represents the height ranges relevant for chemisorbed H atoms. In this region, there is approximately a factor 5 difference in the friction predicted by LDFA and ODF. The lifetime of the vibrational stretch mode of a chemisorbed H-atom at the surface including EHP excitations and lattice vibrations is determined through MDEF simulations for H on Pt(111) adsorbed at the energetically favorable top site using the EMT-PES potential and the respective friction models. A detailed procedure of the simulations is given in the SI and the results are reported in Table~\ref{tab:Lifetimes}.

The vibrational lifetime predicted by MDEF based on LDFA is lower than the result based on ODF and the experimentally measured value. Note that electronic friction in the $^1\mathrm{S}_{0}$ and $^2\mathrm{S}_{1/2}$ states is identical close to the surface and it is, therefore, sufficient to report only one of the two. ODF-based MDEF simulations yield a lifetime that is slightly larger than the experimental value. We expect the prediction to be an upper bound to the experiment as we neglect higher-order phonon-phonon coupling. LDFA overestimates the dissipation rate due to EHP excitations in the chemisorption region, whereas the friction values from ODF are more consistent with experiment.

\begin{table}[]
    \centering
    \begin{tabular}{ccc} \hline\hline
          $\tau_{\bm{q}=0}^\mathrm{LDFA}$ / ps & $\tau_{\bm{q}=0}^\mathrm{ODF}$ / ps  & $\tau^\mathrm{exp}$/ ps\\[2pt] \hline
                    0.174 $\pm$ 0.01  &  1.15 $\pm$ 0.2 & 0.8  \cite{palecek_characterization_2018} \\[2pt] \hline\hline
    \end{tabular}
    \caption{Comparison of experimental and computed vibrational lifetimes of the coherent H-Pt stretch mode of hydrogen at the top site. The theoretical values were acquired with MDEF simulations using LDFA and the $^2\mathrm{S}_{1/2}$ ODF model.}
    \label{tab:Lifetimes}
\end{table}

\begin{figure}[]
    \includegraphics[width=3.3in]{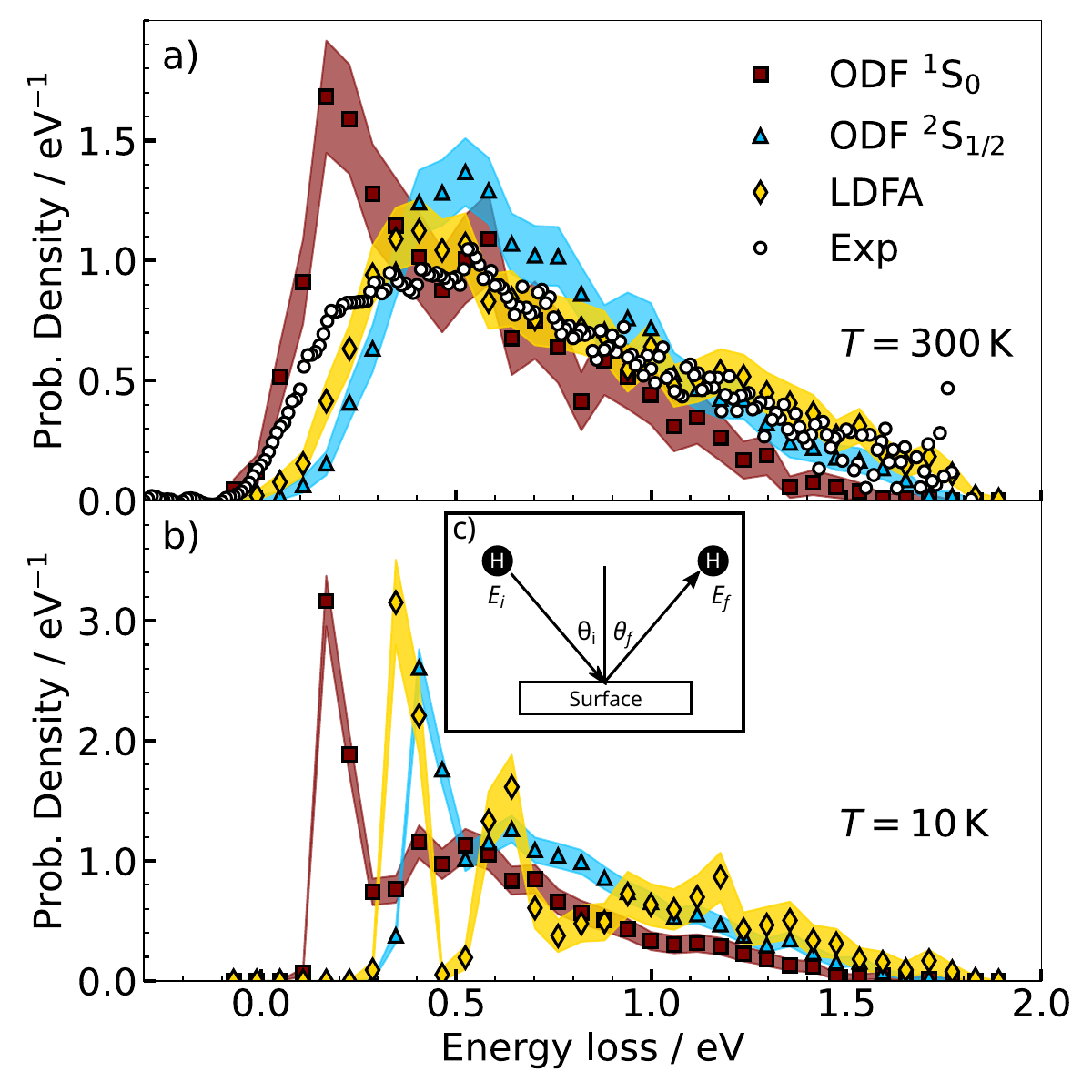}
    \caption{ (a) Computed energy loss distribution obtained from MDEF simulations of specular scattered H atoms from Pt(111) with different friction models. The experimental data is taken from Ref. \citenum{dorenkamp_hydrogen_2018}. The H atoms were launched along the $[10\bar{1}]$ direction with an initial kinetic energy $E_\mathrm{kin,i}$ of 1.92\,eV and an incidence angle of $\theta_\text{i}=45^\circ$. The surface temperature was set to 300\,K.  The coloured, transparent schemes indicate the 95\% confidence interval of the calculated ELDs.  (b) Energy loss distributions simulated at a surface temperature of 10\,K. All other parameters are the same as in panel (a). (c) Inset schematic that depicts the scattering geometry of the simulations.
    }
    \label{fig:ELDs}
\end{figure}

Based on the three different friction models, we simulate hyperthermal ELDs at a kinetic incidence energy of 1.92~eV and a substrate temperature of 300~K. (Figure~\ref{fig:ELDs}) Our LDFA-based MDEF results are consistent with the simulations reported by Dorenkamp \textit{et al.}\cite{dorenkamp_hydrogen_2018} and Lecroart \textit{et al.}\cite{lecroart_adsorbate_2021}, all of which agree well with experiment. The LDFA ELD predicts a lower likelihood of low energy loss scattering than the experiment, but the peak position and tail of the distribution are captured well.  MDEF simulations, performed with the $^1\mathrm{S}_{0}$ ODF model, provide a distribution that is too steep at low energy losses and predicts the most likely energy loss to be around 0.2\,eV significantly below the experimental peak at around 0.5\,eV. Furthermore, scattering events with energy losses beyond 1\,eV are underestimated by this model. Hence, ODF-based MDEF results without inclusion of the spin transition are not consistent with experimental findings. MDEF simulations with the $^2\mathrm{S}_{1/2}$ ODF model predict an energy loss distribution that  manages to qualitatively reproduce the experimental ELD too, but on a quantitative level, LDFA does better. The $^2\mathrm{S}_{1/2}$ ODF curve is able to capture the most probable energy loss around 0.5\,eV seen in the experiment, and also captures the decaying tail at high energy losses correctly within the statistical uncertainties.  

The ELDs predicted by LDFA and ODF $^2\mathrm{S}_{1/2}$ at room temperature are remarkably similar despite their significantly different friction profiles (Figure~\ref{fig:Friction_map}b). 
The random force term in MDEF causes a broadening effect in the predicted room temperature ELDs\cite{hertl_random_2021} which will mask the underlying differences in the description between these two models. For example, previous theoretical works~\cite{Janke_PhD_2016,kandratsenka_unified_2018} have shown the room temperature ELDs predicted by MDEF can be decomposed into strongly overlapping signals from single, double and multibounce events. As such, single scattering events at 300~K cannot be separated clearly from multibounce events. We find that the fraction of adsorbed particles depends on the employed friction model, with LDFA yielding larger sticking coefficients than either ODF model and the $^2$S$_{1/2}$ ODF model predicting higher sticking coefficients than the $^1$S$_{0}$ model (Figure S11). Unfortunately, we are unaware of any reported experimental sticking coefficients for mono-energetic, hyperthermal H atoms impinging on Pt(111) but we propose that such experiments could provide additional support in validating different non-adiabatic simulation techniques.

Furthermore, as previously shown, at low temperatures, energy loss distributions from hydrogen scattering are much more finely resolved and the energy region of single and multiscattering events can be better distinguished.~\cite{hertl_random_2021} ELDs from MDEF simulations at $10$~K (Figure~\ref{fig:ELDs}b) reveal that the differences in the electron friction profiles have a far greater effect at low temperatures. The two ODF models generate similarly shaped ELDs characterised by a distinguishable, sharp first peak and a decaying broad tail, though with a shift to higher energy losses for the $^2\mathrm{S}_{1/2}$ model. It is evident that the electronic friction peak caused by the spin transition introduces a rigid shift between the two ELDs. In contrast, the LDFA model exhibits a clearly separated sharp first peak and a sharp second peak, followed by a broad tail.

\begin{figure}
    \centering
    \includegraphics[width=3.3in]{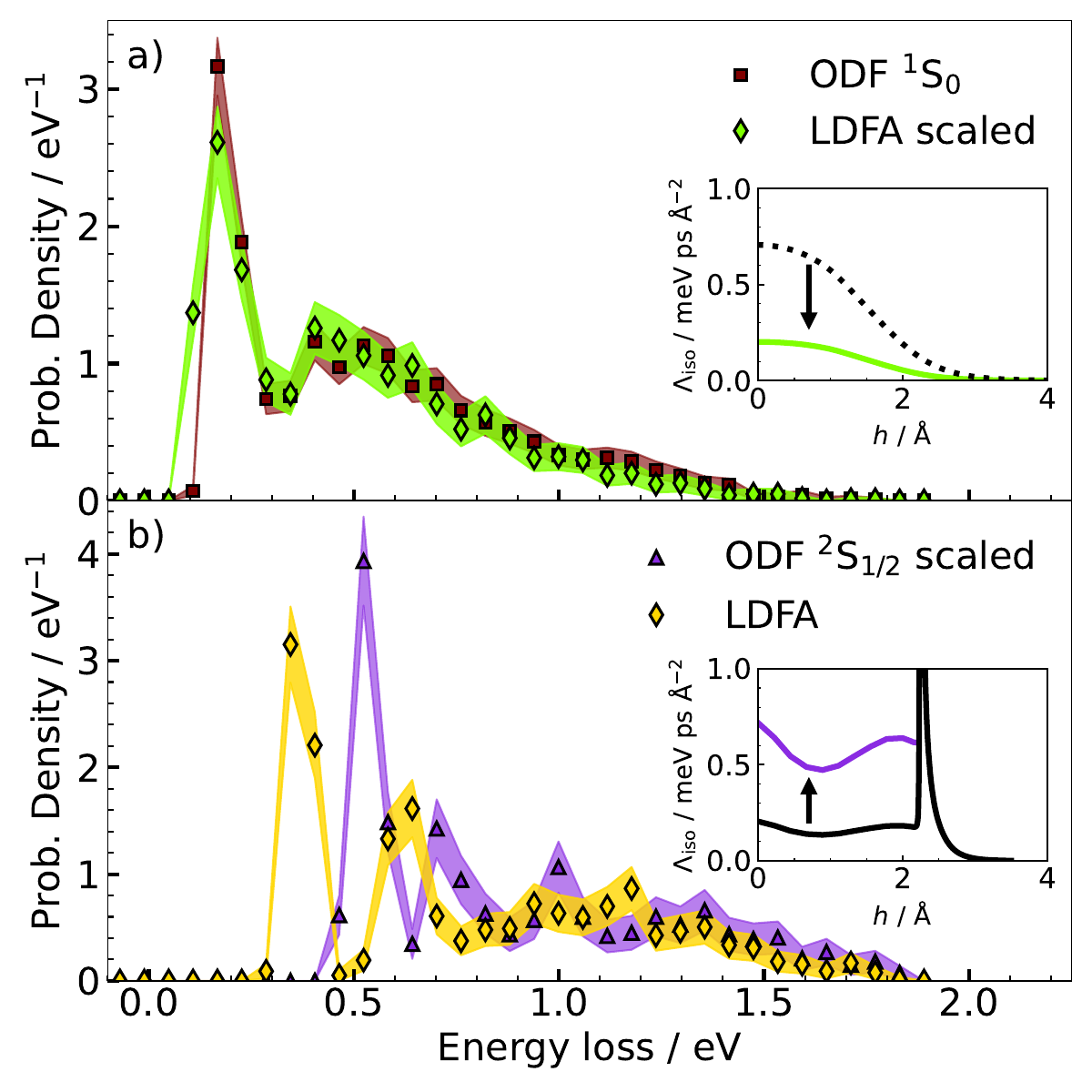}
    \caption{ Comparison of specular energy loss distribution obtained from MDEF simulations with different friction models at a surface temperature of 10\,K. (a) Simulations with a scaled-down LDFA model to match the $^1\mathrm{S}_0$-ODF-based ELD from Figure~\ref{fig:ELDs}. The inset replots \autoref{fig:Friction_map}(b) for the scaled-down LDFA model (green, solid line) and the original LDFA model (black, dotted line). (b) The ELD from MDEF simulations with a scaled-up $^2\mathrm{S}_{1/2}$ model and with the original LDFA model. The inset replots \autoref{fig:Friction_map}(b) for the scaled (purple) and original (black) $^2\mathrm{S}_{1/2}$ ODF-model. The insets are plotted isotropic in direction (i.e the average of diagonal elements, $\Lambda_\mathrm{iso}$)
    }
    \label{fig:ELDs_scaled_friction}
\end{figure}

We examine the relationship between the friction profiles predicted by each model and the low-temperature ELDs by scaling parts of the friction profiles and studying the effects on the simulated ELDs.
Reducing LDFA by a factor of 3.5 provides an ELD (Figure~\ref{fig:ELDs_scaled_friction}a) that closely aligns with that predicted by ODF $^1\mathrm{S}_0$, where the previously distinct second peak and decaying tail have now merged. The rescaled LDFA roughly matches with the average of the diagonal elements ($\Lambda_\mathrm{iso}$) of ODF close to the surface (see insets of Figure~\ref{fig:ELDs_scaled_friction}). 
Similarly, increasing ODF $^2\mathrm{S}_{1/2}$ close to the surface (below the spin transition peak) by the same factor produces a distribution with distinct first and second peaks, akin to LDFA, except with a shift to higher energy losses from the presence of the spin transition peak (Figure~\ref{fig:ELDs_scaled_friction}b). These findings show that by increasing the friction close to the surface, we increase the energetic separation between the peaks in the ELDs at low temperature. Previous analysis\cite{Janke_PhD_2016,hertl_random_2021} connects the distinct peaks in the low-temperature ELD to the number of bounces the H atom undergoes in the trajectory (first peak is due to single bounces, the second peak is due to double bounces, etc.). Thus, by increasing the friction at the surface, there is an increasing translational energy loss that is incurred with each bounce from the surface. This principle is shown for an example trajectory in Supplementary Figure S4, where each bounce accrues the greatest energy loss with the LDFA model.

This study highlights the strengths and limitations of LDFA and ODF for describing nonadiabatic energy loss in high-dimensional hydrogen atom scattering simulations. The LDFA overestimates friction close to the surface compared to electronic friction calculated from first-order response theory based on density functional theory, so-called ODF, and experimental measurements of the vibrational lifetime of the H atom stretch on Pt(111). LDFA furthermore neglects the spin transition that occurs during hydrogen atom impingement. Despite both shortcomings of LDFA, our simulations show that the prevalent room temperature hydrogen atom scattering experiments are well described by simulations with LDFA, closely followed by simulations with ODF where the spin transition is considered in the quasi-adiabatic limit. Therefore, the experiments are not sensitive to even stark and qualitative differences that arise between the two electronic friction profiles. The discrepancy between LDFA and ODF has been previously discussed in literature. Previous work on reactive scattering of N$_2$ on Ru(0001) has shown that ODF and LDFA yield different results, where ODF was found to provide a better description of the experiment.~\cite{spiering_orbital-dependent_2019} The case of H atom scattering is further complicated by the presence of a spin transition that is neither correctly described by LDFA nor ODF and that likely limits the applicability of electronic friction theory.

The specular scattering that is measured corresponds to a highly integrated signal. At room temperature, dynamic fluctuations broaden signatures of single, double and multibounce scattering events, which merge into a single broad distribution. The exact profile of friction determines the energy loss incurred during the initial approach to the surface and the subsequent dynamics at the surface. If friction is large in close vicinity to the surface, single and double bounce events are expected to be energetically well separated in the measured ELDs. If experimentally resolved, this would provide an indirect measurement of the magnitude of electronic friction that would be complementary to vibrational lifetime measurements. At low temperature, details of the energy landscape and the exact profile of friction will become much more apparent in the energy loss distribution. We suggest that low-temperature experiments are conducted in the future, which will support the development of more accurate energy landscapes and help to distinguish signatures of electronic friction related to the chemisorption region and the spin transition during hyperthermal hydrogen scattering. Such experiments will provide deeper insights into the mechanisms of nonadiabatic energy loss during chemical dynamics at metal surfaces and constitute reference results that will further advance the theoretical description of nonadiabatic dynamics at surfaces.

.\section*{Supporting Information.}  Computational details of DFT calculations, simulations, and machine learning models, as well as additional simulation data.

\section*{Note} The authors declare no competing financial interests.

\section*{Acknowledgments}

\bibliography{mylib}

\end{document}


\tableofcontents



\section{DFT calculations}
    
Our DFT calculations were performed with the FHI-aims code (2021
development version).\cite{blum_ab_2009} To acquire the necessary input data for the ODF model from DFT, we used a $2\times2$ cell with 6 layers with a single H atom in the cell to model atomic hydrogen interacting with a Pt(111) surface. The two bottom layers were kept frozen in their bulk truncated positions. Periodic boundary conditions were employed in all three directions. The cell height (perpendicular to the slab) was 70\,\,{\AA}, to prevent the interaction of the slab with its periodic images in the vertical direction. The lattice constant was taken from Ref.\citenum{lecroart_adsorbate_2021}. For the relaxation of the slab, we used a standard 'tight' basis set. The reciprocal space was sampled with a $16 \times 16 \times 1$ $\bm{k}$-point mesh according to the sampling method proposed by Monkhorst and Pack,\cite{monkhorst76} and an atomic ZORA relativistic correction.\cite{ZORA} A Gaussian width of 0.1\,eV was used for the occupation smearing in the self-consistent field cycles. The PBE exchange-correlation functional\cite{pbe1, pbe2} is employed with $10^{-6}$~eV, $10^{-3}$~eV and $10^{-5}$~e/$\mathrm{a}_0^3$ tolerances for the total energy, the eigenvalue energies and the electronic density, respectively. The convergence criterion for the forces was set to $10^{-4}\,\mathrm{eV}/${\AA}.


%

The ODF tensors were calculated using FHI-aims,\cite{blum_ab_2009} and we refer to Ref. \citenum{box_ab_2023} for a detailed description of the underlying formulae of ODF implemented in that electron structure code.
The same computational settings were employed as above, except a standard FHI-aims `light' numerical basis set was used. 
To assess the effect of the spin on the friction tensor, we conducted calculations with and without spin polarisation and kept the total spin of the system fixed. 
We used a Gaussian width, $\sigma$, of 0.6 eV to replace the delta distribution in the expression of the ODF friction tensor. The ODF calculations are treated at an electronic temperature of 0~K.

\begin{figure}
    \centering
    \includegraphics[width=0.8\textwidth]{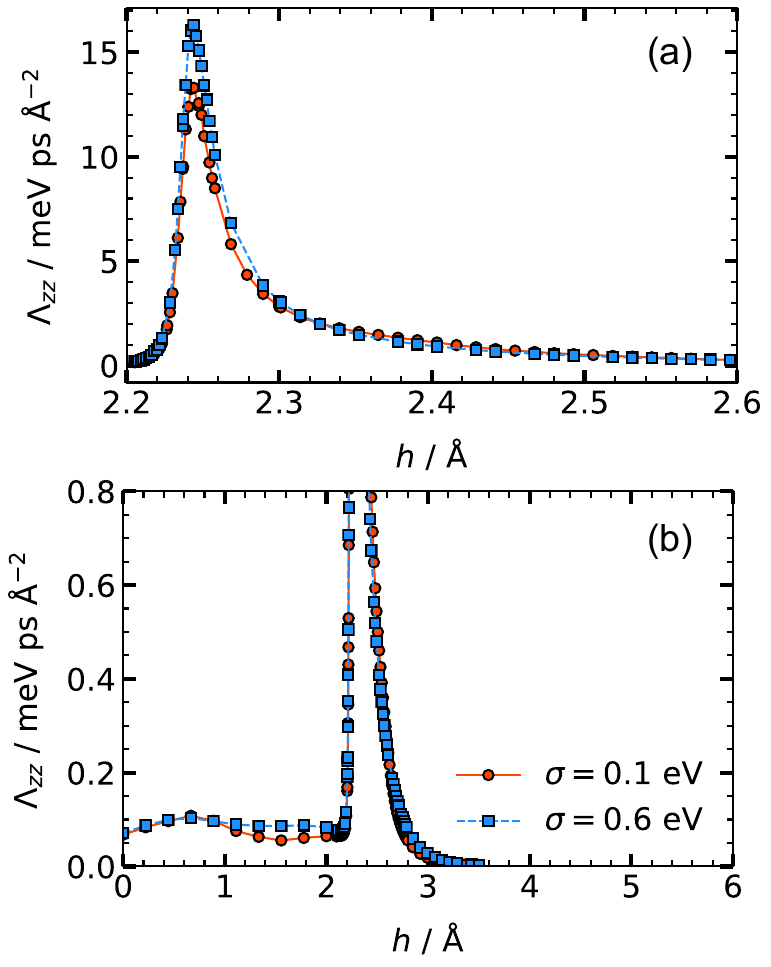}
    \caption{Same as Figure 1(b) in the main manuscript, except only $^2\text{S}_{1/2}$ ODF is shown for two different broadening parameters, $\sigma$. The two subplots are shown for different axis limits.}
    \label{fig:friction-sigma}
\end{figure}

The spin-polarised ODF coefficients are not significantly different between $\sigma=0.1$~eV and $\sigma=0.6$~eV for a range of heights above the hcp site (as shown in \autoref{fig:friction-sigma}) except in the range of heights $1-2$~$\mathrm{\AA{}}$ and at the maximum of the spin-transition peak. These differences are not expected to strongly affect the energy loss distributions nor affect the conclusions drawn in the main paper. 

\begin{figure}
    \centering
    \includegraphics[width=0.9\textwidth]{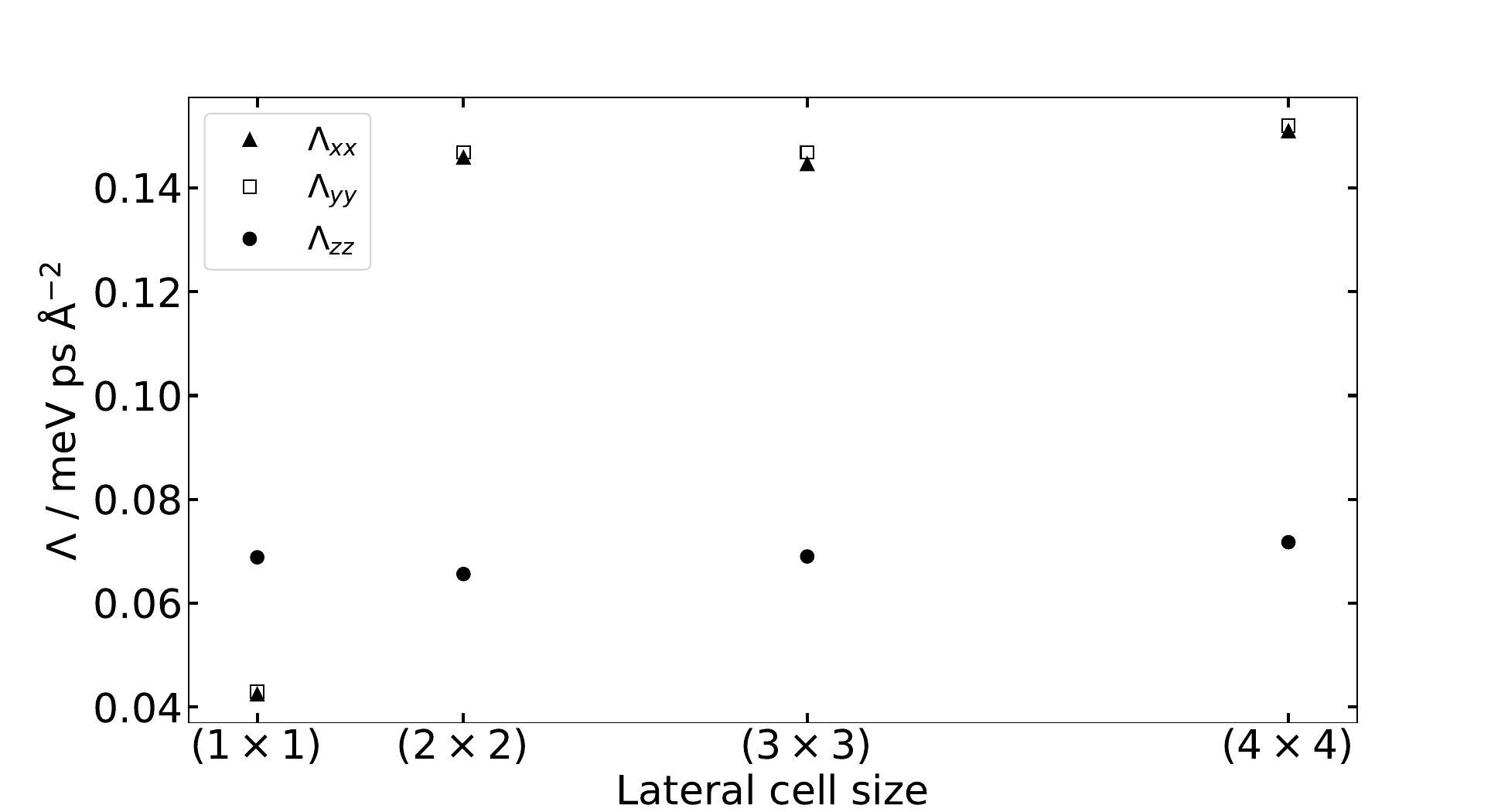}
    \caption{Diagonal elements of the ODF friction tensors as a function of the lateral cell size for a H atom in its minimum energy configuration at the fcc-hollow site, depicted in Figure\,1a).}
    \label{fig:Friction_size}
\end{figure}

We also assessed the convergence with respect to the lateral cell size and found that the diagonal elements are converged in a $p(2\times2)$ cell as shown in Figure\,\ref{fig:Friction_size}.


\section{Simulation details}
The MDEF simulations were carried out using the NQCDynamics.jl\cite{gardner_nqcdynamics_2022} package (development version based on version 0.12) that interfaces with the md\_tian2 code,\cite{MDT2GIT} which provides the EMT-PES for the energies, and forces required for the propagation of the atoms. The parameters that define the H/Pt(111) EMT-PES have been taken from Ref.\citenum{kammler_genetic_2017}.  The BAOAB Langevin propagator proposed by Leimkuhler \textit{et al.} \cite{leimkuhler_robust_2013} has been used for all atoms in the simulation cell except that we only provide non-zero friction elements for the H atom to account for ehp excitation during the scattering process. Slab atom positions were randomly sampled with thermal Monte Carlo simulations, which were resampled for every 100 scattering trajectories. Velocities for the slab atoms were obtained from a Boltzmann distribution. Incidence kinetic energy, incidence angle and crystallographic incidence direction were given as initial conditions, which define the initial entries of the H atom's velocity vector. The initial lateral positions of the H atom were chosen randomly and the projectile was positioned 6\,{\AA} above the surface. The simulation time and the time step were chosen manually for the chosen friction method and initial kinetic energy. Table \ref{tab:hatom_sim_details} gives an overview of the simulation details.
\begin{table*}
    \centering
    \caption{Chosen MDEF simulation parameters.} \label{tab:hatom_sim_details}
    \begin{tabular}{lcccc} 
    \hline
     Model & $E_\mathrm{kin,i}$ & Time step / fs & Simul. time limit / fs \\
     \hline



     LDFA & 1.92 & 0.1 & 300 \\
     ODF $^1\text{S}_0$ & 1.92 & 0.1 & 300 \\
     ODF $^2\text{S}_{1/2}$ & 1.92 & 0.05 & 300 \\
    


     \hline
    \end{tabular}
\end{table*}
To meet the experimental conditions as close as possible, we only used those scattering events for the calculation of the energy loss distribution, which scattered in-plane with a tolerance angle of $6^\circ$ at a scattering angle of 45$\pm5^\circ$. The shaded regions of Figure 2 in the main manuscript give the 95 \% confidence intervals, $C$ which are calculated by the following:
\begin{equation}
    C = \frac{1.96}{w} \times \sqrt{\frac{p(1-p)}{N}},
\end{equation}
where $w$ is the bin width of the ELDs, $p$ is the probability density and $N$ is the number of trajectories.

\section{Friction models}
Within the LDFA framework, the friction tensor is isotropic and characterized by a single scalar, $\bm{\Lambda}=\Lambda \mathbb{1}$, where $\Lambda$ is determined by the electron density. A notable advantage of the EMT approach is that the total energy expression depends on a model density, $n_\mathrm{EMT}$, which can be used to compute the friction coefficients at the LDFA level \cite{janke_accurate_2015, hertl_electronically_2022}. We employed the mapping function between model density and friction coefficient provided in md\_tian2, as detailed in Ref.\citenum{janke_accurate_2015}.


When ODF coefficients were employed, they came from one of two Gaussian process regression\cite{deringer_gaussian_2021} models: one without spin polarisation and one with spin polarisation combined with a 1D fit for H atom heights above the beginning of the spin transition.  
The GPR~\cite{rasmussen_gaussian_2005} models were generated using \texttt{scikit-learn}~\cite{pedregosa_scikit-learn_2011} (version 1.2.0) with a vector of the inverse hydrogen atom distances to the platinum atoms used as a descriptor. 

$3198$ data points were collected via spin-unpolarised DFT calculations, and the $757$ structures with the hydrogen atom above $1.8$~$\Angstrom$ from the surface were recalculated with spin-poloarisation to form the two datasets. 
Of the data points, $95$~\% were chosen to be training data, the remaining $5$~\% formed the test data.
As a consequence of the exclusion of configurations where the hydrogen atom sits below the second subsurface platinum layer in the training data, we do not employ the generated ODF models for incidence energies where significant subsurface scattering is expected. 
The \texttt{scikit-learn} convenience function, \texttt{MinMaxScaler}, is employed on the training and test data to scale the features to a range between $0$ and $1$.

\begin{figure}
    \centering
    \includegraphics[width=0.9\linewidth]{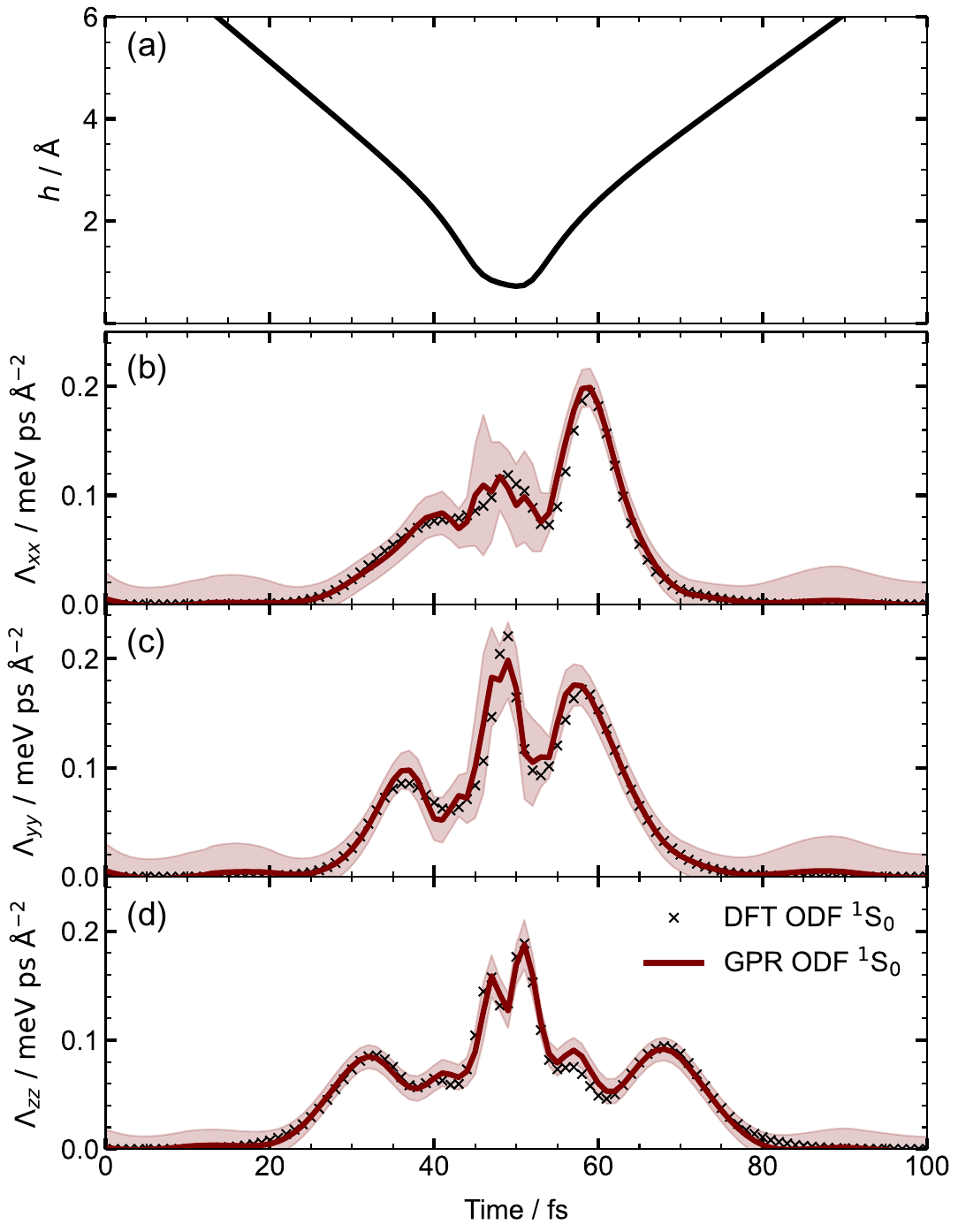}
    \caption{For an example trajectory ($E_\mathrm{i}=1.92$~eV) the (a) height of the hydrogen atom projectile and (b-d) diagonal ODF $^1\mathrm{S}_0$ friction tensor elements calculated with spin unpolarised DFT and the GPR models are shown as a function of simulation time. 
    Trajectory is unseen in the training of the GPR models. 
    Shaded region is the standard deviation provided by the  GPR model.}
    \label{fig:s0_validation_traj}
\end{figure}

Only the diagonal elements of the electronic friction tensor (EFT) were included in the GPR models and the subsequent MDEF simulations.
The   $^1\mathrm{S}_0$-ODF model achieved a RMSE of $0.0076$~$\ips$ for the diagonal elements of the EFT for $5$~unseen validation trajectories, an example of which is shown in \autoref{fig:s0_validation_traj} 
The GPR models quantitatively predict the $^1\mathrm{S}_0$-ODF diagonal elements across this example trajectory, with the accuracy highest and GPR standard deviation lowest for the $z$ diagonal element. 

\begin{figure}
    \centering
    \includegraphics[width=0.9\linewidth]{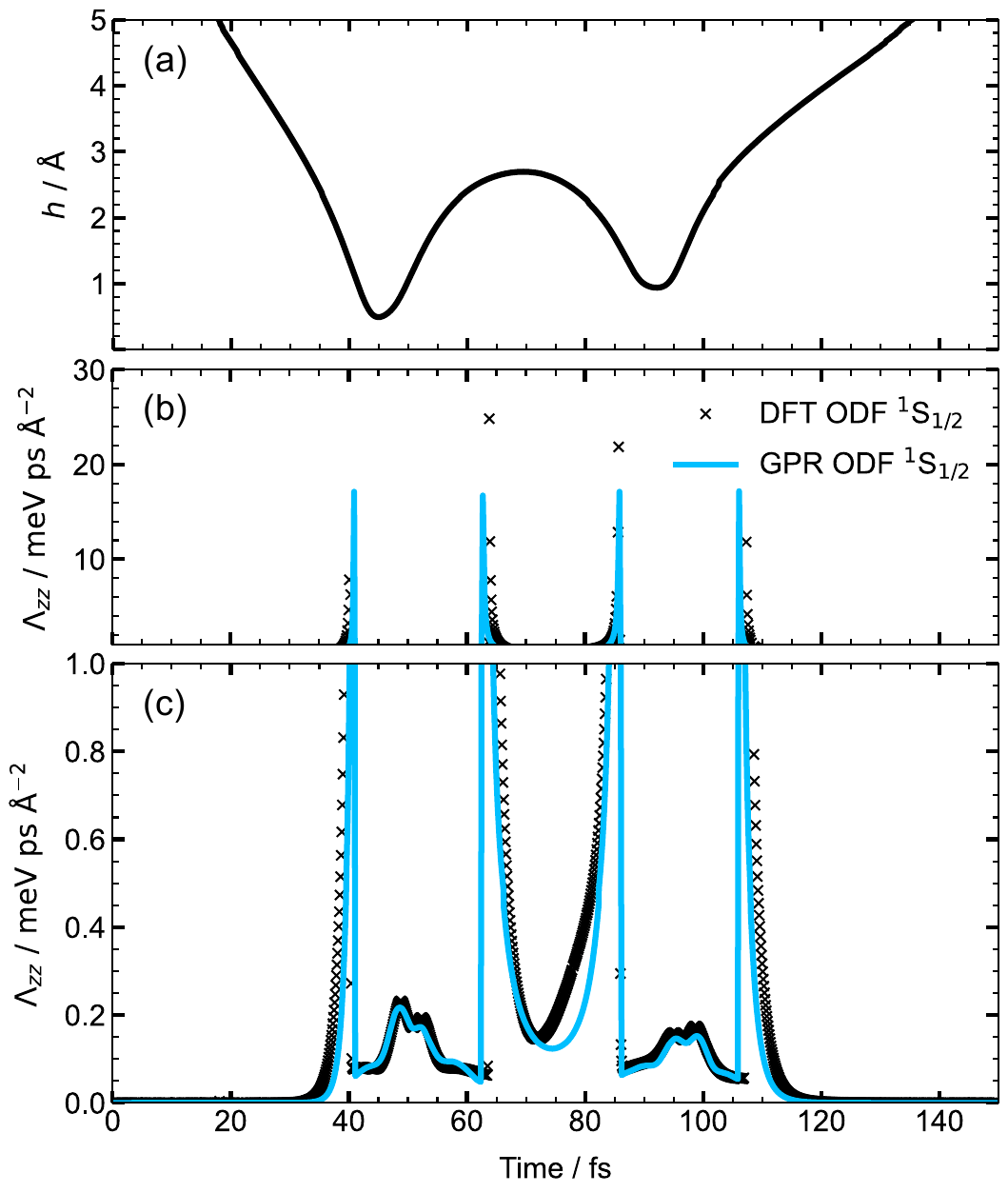}
    \caption{For an example trajectory ($E_\mathrm{i}=1.92$~eV), the (a) height of the hydrogen atom projectile and (b-c) ODF $^2\mathrm{S}_{1/2}$ $\Lambda_{zz}$ element calculated with spin-polarised DFT and GPR model are shown as a function of simulation time.
    Trajectory is unseen in the training of the GPR model. 
    Different $y$-axis limits are shown in (b) and (c) for visual clarity.}
    \label{fig:s1_validation}
\end{figure}

Direct fitting of the $^2\mathrm{S}_{1/2}$-ODF coefficients using the same methodology was not possible, due to the presence of the large peak in diagonal element associated with motion in $z$ coordinate ($\Lambda_{zz}$), as a consequence of the screening of the projectile spin, as discussed in the main manuscript.
Instead, a GPR model was generated for the ODF coefficients below $2.21$~$\Angstrom$ and the mass-weighted friction ($\tilde{\Lambda}$) in the spin-transition region was fitted separately using:
\begin{equation} \label{eq:spinfit_curves}
    \tilde{\Lambda}_{zz}(h)=\left\{\begin{array}{cc}
p(h,h_a,h_0,h_\omega,h_\sigma), & 2.21\ \Angstrom \leq h \leq 2.24\ \Angstrom \\
p(h,h'_a,h'_0,h'_\omega,h'_\sigma), & 2.24\ \Angstrom < h \leq 2.5\ \Angstrom \\
g(h,h''_a, h''_t, h''_\omega, h''_\sigma), & h>2.5\ \Angstrom
\end{array}\right. ,
\end{equation}
where the Pearson type VII-like distribution~\cite{pearson_x_1895}, $p$ was used as a functional form:
\begin{equation}
    p(h,h_a,h_0,h_\omega,h_\sigma) = \frac{h_a}{\left[1+\left[h^{-1}_
    \sigma\left(2^{1 /h_\omega}-1\right)^{\frac{1}{2}}\left(2h - 2h_0\right)\right]^2\right]^{h_\omega}}, 
\end{equation}
and the exponential decay ($g$) function is given by:
\begin{equation}
    g(h, h_a, h_0, h_\omega, h_\sigma) = h_a * \exp(-h_0 * h + h_\sigma) + h_\omega .
\end{equation}
The employed parameters are given in \autoref{tab:parameters}.
\begin{table}[]
    \centering
    \caption{Parameters employed for the curve fitting of the spin-peak in the $^2\mathrm{S}_{1/2}$-ODF coefficients as employed in \autoref{eq:spinfit_curves}}
\begin{tabular}{cccccc}
\hline
\hline
    $h_a$ & $1.0005$ $\ips$ & $h^{\prime}_a$ & $1.0546$ $\ips$ & $h^{\prime \prime}_a$ & $13.1754$ $\ips$ \\
    $h_0$ & $2.2430$ $\Angstrom$ & $h^{\prime}_0$ & $2.2397$ $\Angstrom$ & $h^{\prime \prime}_0$ & $6.0807$ $\Angstrom^{-1}$ \\
    $h_\omega$ & $2.8186$ & $h^{\prime}_\omega$ & $0.6428$ & $h^{\prime \prime}_\omega$ & $0.0000$ $\ips$ \\
    $h_\sigma$ & $0.0175$ $\Angstrom$ & $h^{\prime}_\sigma$ & $0.0452$ $\Angstrom$ & $h^{\prime \prime}_\sigma$ & $78.9657$ \\
\hline
\end{tabular}
\label{tab:parameters}
\end{table}
The fitted curve described above is used for the $\Lambda_{zz}$ coefficients when the hydrogen atom is above $2.21$~$\Angstrom$ whilst the GPR model is used everywhere else.
The performance of this model for $\Lambda_{zz}$, shown for an example trajectory in \autoref{fig:s1_validation}, is adequate to qualitatively explore how a peak in friction in the spin transition region affects the ELDs (see the main manuscript).

\section{Energy loss profile}
\begin{figure}
    \centering
    \includegraphics[width=0.9\textwidth]{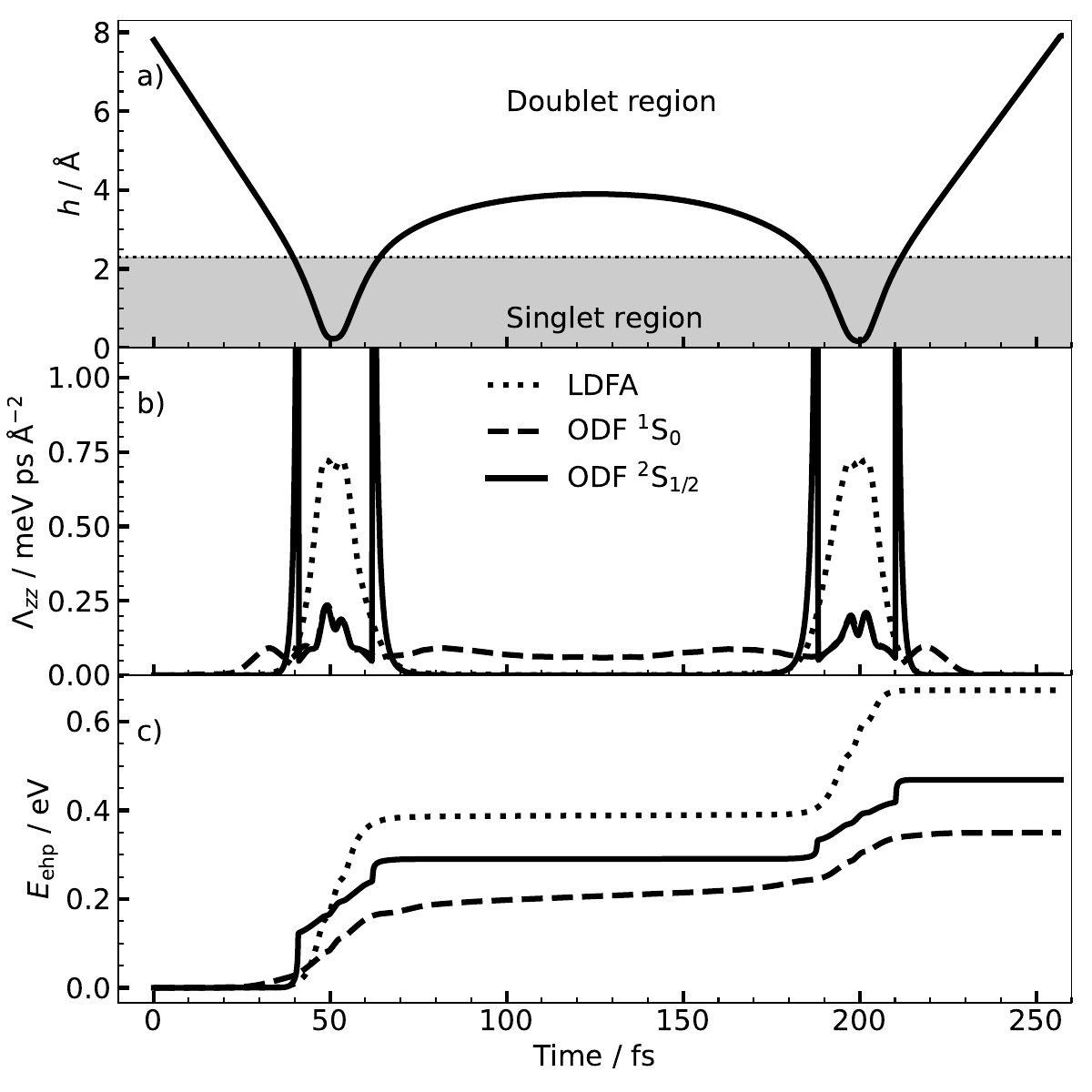}
    \caption{Altitude, friction and non-adiabatic energy loss profile acquired from the positions and velocities of an adiabatic trajectory with two collision events. Panel a) depicts the altitude of the projectile, where $h=0$\,{\AA} corresponds to the surface layer. The grey-shaded area corresponds to the region where the H atom's spin is screened by the metal electrons{---}see Figure\,1a) for details. Panel b) shows the friction profile for the three friction models along the time evolution of the scattering trajectory. Panel c) shows the corresponding accumulated energy loss over time. }
    \label{fig:eloss-profile}
\end{figure}

In this section, we provide details on how the energy loss profile along the trajectory in Figure\,\ref{fig:eloss-profile} was acquired. At first, a Born-Oppenheimer molecular dynamics (BOMD) trajectory was calculated without non-adiabatic effects but all other conditions are the same as those used in the simulations for the ELDs shown in Figure\,2. The H atom's position was used to compute the friction tensor which the projectile would face in a MDEF trajectory at that location. The extracted friction tensor and velocities were subsequently used to calculate the energy transfer due to electronic friction along the trajectory with the following equation:
\begin{equation}
    E_\text{ehp} = \int_0^t \text{d}t' \dot{\vec{r}} \bm{\Lambda}(\vec{r}) \dot{\vec{r}} \approx \sum_{i=1}^N \dot{\vec{r}}_i \bm{\Lambda}(\vec{r}_i) \dot{\vec{r}}_i \Delta t,
\end{equation}
where $\bm{\Lambda}(r)$ and $\dot{\vec{r}}$ are the position-dependent friction tensor and the projectile's velocity, respectively. The index $i$ marks the $i$th integration step and $N$ stands for the trajectory's total number of integration steps. The velocities taken from the BOMD trajectory were not affected by the frictional drag during the run and are thus higher compared to an actual MDEF trajectory. Hence, for each degree of freedom, we subtract the accumulated energy loss along that respective degree of freedom from the magnitude of the adiabatic velocity component $|\dot{r}_{\alpha, j}^\mathrm{BOMD}|$, i.e.,
\begin{equation}
    |\dot{r}_{\alpha, j}| =  |\dot{r}_{\alpha, j}^\mathrm{BOMD}| -  \sqrt{2\frac{\sum_{i=1}^{j-1} \Lambda_{\alpha\alpha}(\vec{r}_i) \dot{r}_{\alpha,i}^2 \Delta t}{m_\mathrm{H}}}.
    \label{eq:velo_correction}
\end{equation}
Here $\alpha$ labels the Cartesian elements and $j$ stands for the $j$th integration step of the BOMD trajectory. The sum in the second term of Eq.~\ref{eq:velo_correction} runs over all previous integration steps. 

For the example trajectory in Figure\,\ref{fig:eloss-profile} a) one can infer from the H atom height that there are two bounces on the Pt(111) surface, the first time around 50\,fs and the second time around 200\,fs. 
Hence, along this trajectory, the particle undergoes a spin transition four times in the $\Lambda_{zz}$ profile of the $^2\mathrm{S}_{1/2}$-ODF model, shown in Figure\,\ref{fig:eloss-profile}b) where the friction reaches its maximum values. In contrast the maxima of the LDFA can be found when the H atom is closest to the Pt surface. The $^1\mathrm{S}_{0}$-ODF model provides significant friction at much larger distances from the surface, which is a consequence of the H atom's wrong spin state at those distances. 
These differences have an effect on the time profile of the energy dissipation in Figure\,\ref{fig:eloss-profile} c).
In the case of LDFA, steep increases in energy dissipation are localised to the two bounces with the surface. 
The $^1\mathrm{S}_{0}$-ODF model exhibits lower energy dissipation than LDFA at the bounces, but continually dissipates energy far from the surface (in the time spent between bounces).  
In the case of the $^2\mathrm{S}_{1/2}$-ODF model, the energy dissipation exhibits steep jumps when the H atom passes through the spin transition region, with a lower rate of energy dissipation close to the surface than LDFA. Note that the first pass through the spin transition regions makes the largest contribution of all four passes, where the velocity is the highest. 
In this example trajectory, the difference in energy loss between the first and second bounce is greatest for LDFA, and similar for both ODF models. 
This is expected to account for the distinction in the peaks associated with single bounce and double bounce trajectories in the low-temperatue ELDs for LDFA compared to the ODF models. 

\section{The role of non-diagonal elements of EFT}

\begin{figure}
    \centering
    \includegraphics[width=0.8\linewidth]{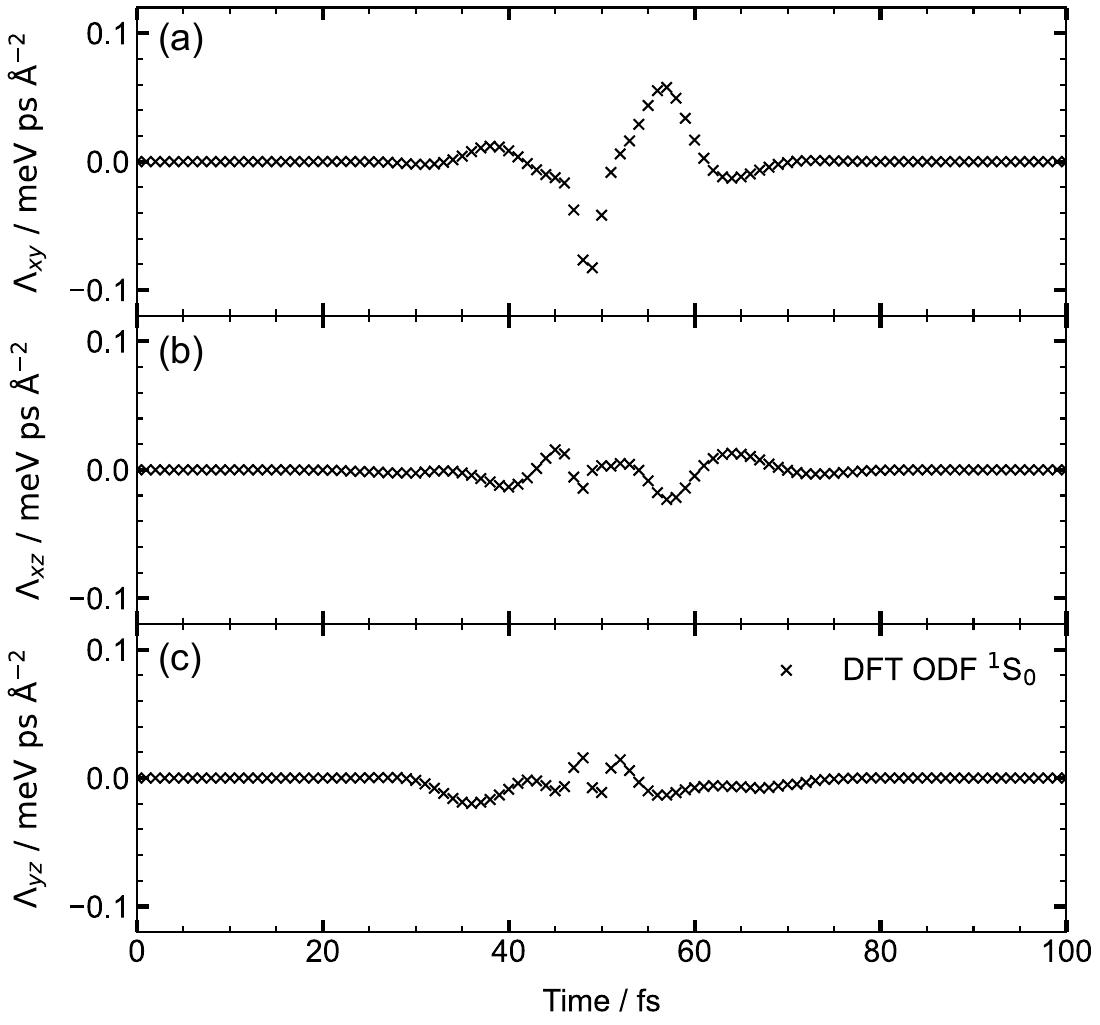}
    \caption{For an example trajectory ($E_\mathrm{i}=1.92$~eV) the (a-c) off-diagonal ODF $^1\mathrm{S}_0$ friction tensor elements.}
    \label{fig:off-diagonal-traj}
\end{figure}

The off-diagonal ODF elements were excluded from the GPR models used for the results in the main manuscript. The off-diagonal elements have a small magnitude except for the closest distances between the projectile, shown for an example trajectory in \autoref{fig:off-diagonal-traj}. At these distances, the hydrogen atom velocity is small, and the subsequent friction-induced energy loss is expected to be negligible. The off-diagonal elements are thus not expected to contribute significantly to the energy loss distributions.

\begin{figure}
    \centering
    \includegraphics[width=0.5\linewidth]{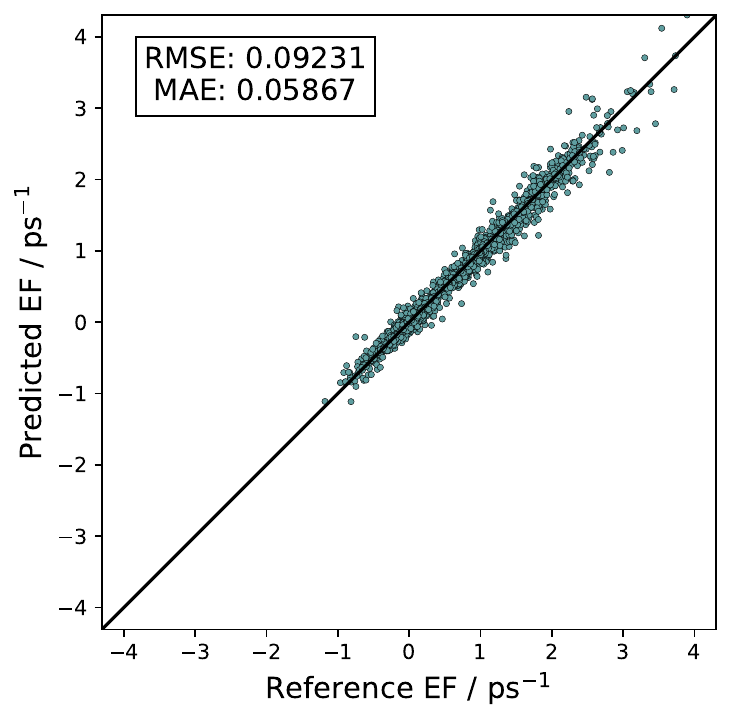}
    \caption{Performance of ACE-based EFT model based on ODF. Predictions of mass-weighted electronic friction tensor elements in ps$^{-1}$ made using an ACE-based model with corresponding DFT-based reference results. All diagonal and off-diagonal elements are compared.}
    \label{fig:ace-eft-model}
\end{figure}

To assess the impact of off-diagonal elements of EFT on the scattering dynamics, we constructed a model based on a method by Sachs~\textit{et~al.}\cite{sachs_equivariant_2024}. This approach utilizes the atomic cluster expansion (ACE)\cite{drautz_atomic_2019} methodology to construct full EFTs. The model was trained using Adam optimizer, with a learning rate of 10$^{-4}$, and exponential decay rate for the momentum and velocity terms of 0.99, and 0.999, respectively. A column-wise coupling was used in constructing a representation of atoms together with a cutoff distance of 5~\AA, a maximum correlation order of 2, and a maximum polynomial degree of 12. Scalar- and matrix-equivariant sub-blocks of the diffusion coefficient matrix were included.
The model RMSE evaluated on a test set consisting of 634 random H/Pt structures was 0.092~ps$^{-1}$ (Fig.~\ref{fig:ace-eft-model}).

\begin{figure}
    \centering
    \includegraphics[width=0.9\textwidth]{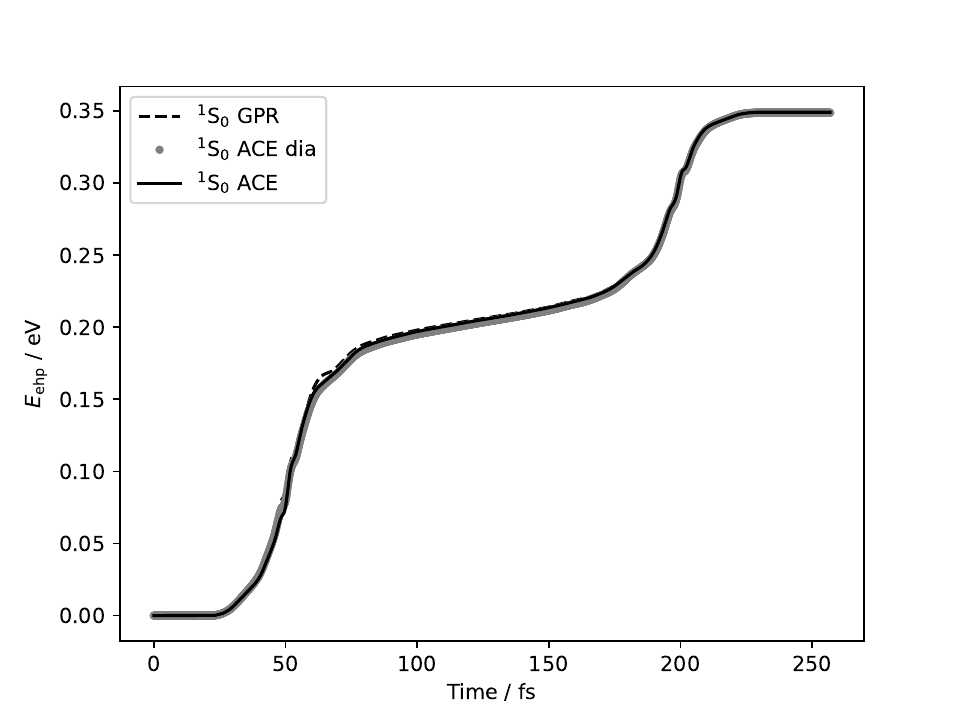}
    \caption{Accumulated energy loss over time (analogous to Fig.~\ref{fig:eloss-profile}c) evaluated using the GPR model (dashed line), ACE model with diagonal elements (grey circles), and ACE model with all the elements of EFT (solid line).}
    \label{fig:eloss-ace}
\end{figure}

\begin{figure}
    \centering
    \includegraphics[width=0.8\textwidth]{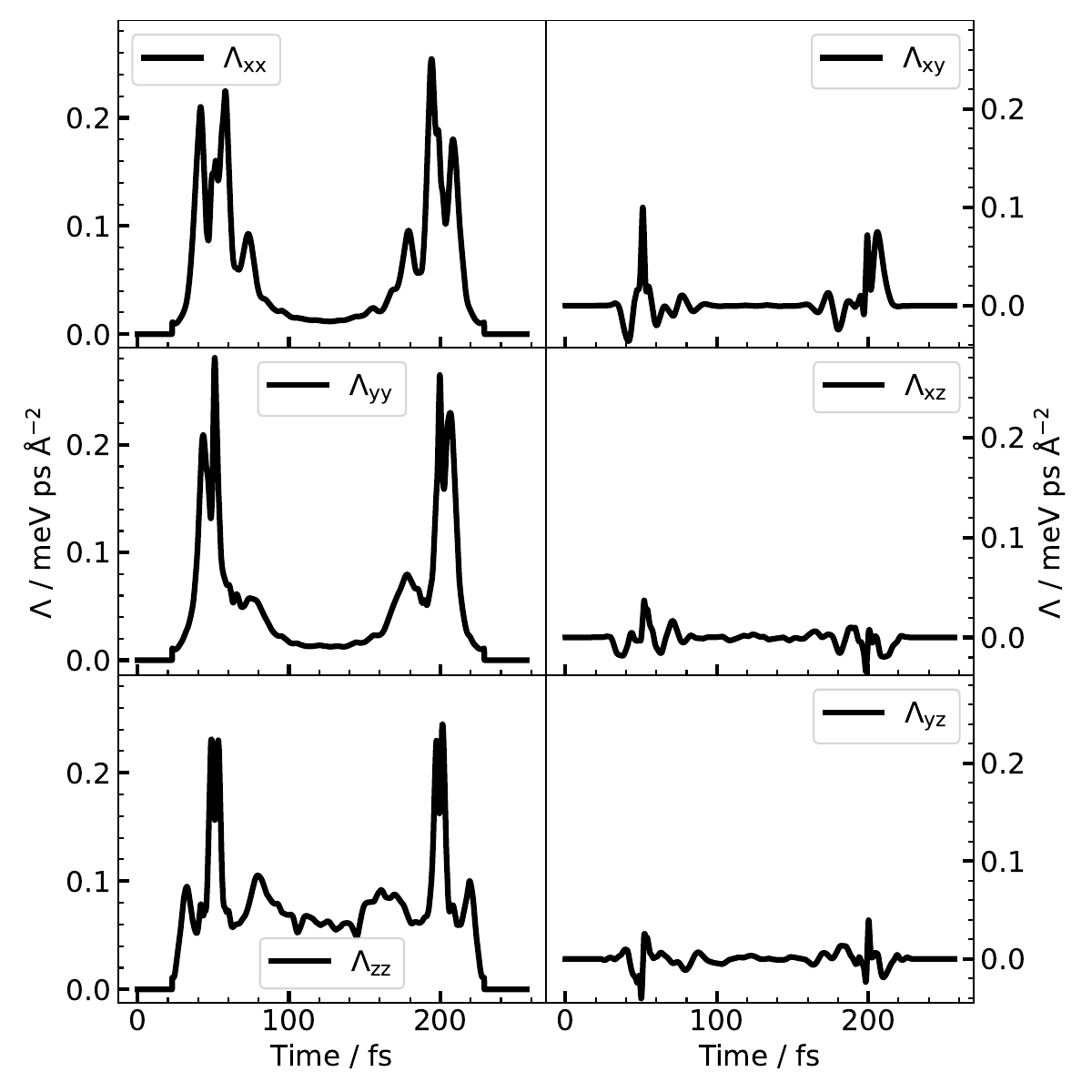}
    \caption{Different EFT elements along the scattering trajectory (same as in Fig.~\ref{fig:eloss-profile}c) evaluated using the $^1S_0$ ACE model.}
    \label{fig:eft-elems-ace}
\end{figure}

In Fig.~\ref{fig:eloss-ace}, the energy loss is shown for the same scattering trajectory as in Fig.~\ref{fig:eloss-profile} c. Here, the energy loss obtained using the GPR ODF model is compared with the energy loss profile generated using the ACE model in two versions. In the first version, only diagonal elements were used, whereas in the second version, all the elements of EFT were included. The energy dissipation generated with the ACE model follows the curve generated with the GPR model very closely. Similarly, the agreement between the diagonal-only and full tensorial ACE models is almost perfect. Nondiagonal EFT elements are roughly 10 times lower than the diagonal elements, throughout an entire scattering trajectory (Fig.~\ref{fig:eft-elems-ace}). This suggests that the impact of nondiagonal elements of EFT on the scattering dynamics of H at Pt(111) surface is minor and thus neglecting the nondiagonal elements is justified.

\section{Calculation details of dynamical lifetimes}
\begin{figure}[]
    \includegraphics[width=1\textwidth]{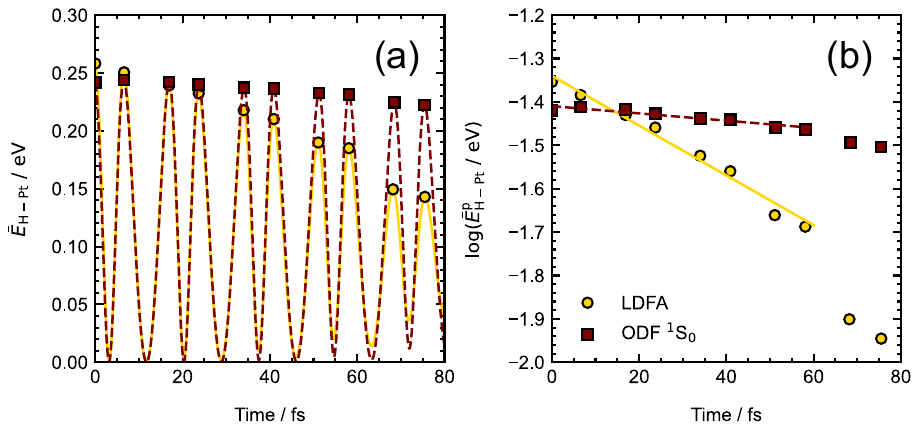}
    \caption{ Average kinetic energy of the H-Pt symmetric stretch vibration, $\bar{E}_{\text{H-Pt}}$, plotted as a function of time.
    The dynamics are simulated by the EMT-LDFA and $^1\mathrm{S}_{0}$ ODF models and the energy-time profile is plotted in panel (a). The points in panel (a) mark the amplitudes $\bar{E}_{\text{H-Pt}}^{\text{P}}$ which are used for the fit to extract the dynamical lifetime, shown in panel (b). Data points after 60\,fs were excluded from the fit because the H atom would typically switch adsorption sites at this time.}
    \label{fig:lifetimes}
\end{figure}
In order to obtain vibrational lifetimes that include anharmonic effects, we employed MDEF simulations where the H atom in an adsorption geometry at a top site is initiated with kinetic energy equal to the experimental\cite{palecek_characterization_2018} vibrational energy for $v=1$ (0.259~eV). These simulations follow the same procedure to initiate the metal atoms as the scattering simulations described above.
Subsequently, we let the system propagate with a time-step $\Delta t = 0.1$\,fs  and monitored the kinetic energy decay along the H-Pt bond. A total of 1,000 simulations were performed for each LDFA and ODF, and the energy loss was averaged over those trajectories. Figure \ref{fig:lifetimes} shows the resulting energy-time profile. A decaying exponential $\bar{E}_{\text{H-Pt}} = \bar{E}_{\text{H-Pt}}^{\text{P}} e^{-kt}$ was fitted to the amplitudes, the inverse of the constant $k$ corresponds to the lifetime $\tau_{\bm{q}=0}^{\text{i}}$, where $\text{i}\in \{\text{LDFA}, \text{ODF} \}$. We discarded data after 60\,fs because the H atom would typically diffuse to neighbouring sites at this point.
Table\ref{tab:Lifetimes_methods} compares the dynamic lifetimes detailed above to vibrational lifetimes calculated via the harmonic approximation using finite displacements{---}see Ref.\citenum{maurer_ab_2016} for details. This approach neglects anharmonic effects, but, in the case of ODF, we can evaluate friction at the perturbing frequency of the vibration and therefore do not have to make the zero-frequency (quasi-static) approximation as in MDEF. The corresponding ODF vibrational lifetime value therefore accounts for the quantum nature of the vibration (in the harmonic limit). 
\begin{table}[]
    \centering
    \begin{tabular}{cccc} \hline\hline
         Method & $\tau_{\bm{q}=0}^\mathrm{LDFA}$ / ps & $\tau_{\bm{q}=0}^\mathrm{ODF}$ / ps & $\tau^\mathrm{exp}$/ ps \\[2pt] \hline
         Dynamic &  0.17 $\pm$ 0.01  &  1.15 $\pm$ 0.2 & - \\[2pt] \hline
         Static-Harmonic  & 0.19 & 1.11 & - \\ [2pt] \hline
         Exp. & - & - & 0.8 \cite{palecek_characterization_2018} \\[2pt] \hline
    \end{tabular}
    \caption{Comparison of vibrational lifetimes acquired dynamically via MDEF and statically via DFT-based electronic structure calculations.}
    \label{tab:Lifetimes_methods}
\end{table}
Based on the excellent agreement between the vibrational lifetimes obtained via static and the dynamic methodology, we can conclude that the classical treatment of the hydrogen motion does not seem to be important for the determination of dynamical lifetimes for chemisorbed hydrogen on metal surfaces at room temperature. 
\section{Sticking coefficients}
\begin{figure}[]
    \includegraphics[width=0.65\textwidth]{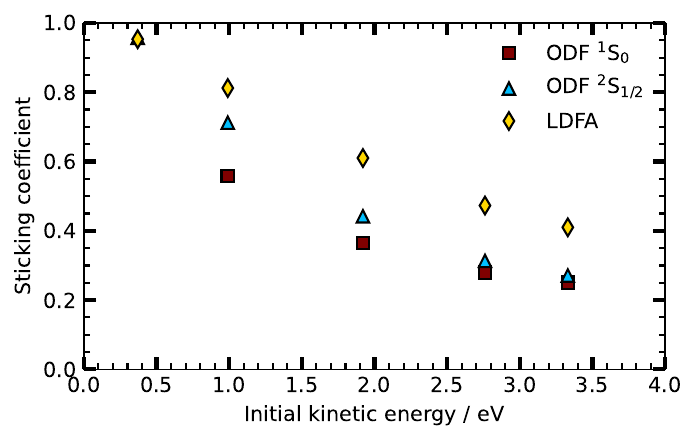}
    \caption{ Sticking coefficients as a function of the initial kinetic energy from MDEF simulations with different friction models.  The initial incidence angle is $\theta_\text{i}=45^\circ$ and the H atoms lateral incidence direction was set to the $[10\bar{1}]$ direction. All initial conditions were chosen to be as close as possible to the experiment described in Ref. \citenum{dorenkamp_hydrogen_2018}.}
    \label{fig:Sticking}
\end{figure}
It is worthwhile to investigate how different friction theories affect the likelihood for adsorption. We launched MDEF simulations with the three friction models and varied the initial kinetic energy $E_\text{kin,i}$. The fraction of adsorbed atoms are depicted in Figure\,\ref{fig:Sticking}. For $E_\text{kin,i}$=0.37\,eV, almost all H atoms remain at the surface. For the remaining initial kinetic energies, our LDFA values are in line with those reported for Ni(111), Ag(111) and Au(111).\cite{dorenkamp_hydrogen_2018} Additionally, LDFA always gives larger sticking coefficients than ODF. This is consistent with the observation made by Spiering and Meyer for dissociative adsorption of H$_2$/Cu(111)\cite{spiering_testing_2018} and N$_2$/Ru(0001)\cite{spiering_orbital-dependent_2019}, albeit the differences between LDFA and ODF are more pronounced for our methodology. Furthermore, we note that the differences between the singlet and the doublet ODF model get smaller with increasing kinetic energy.  Unfortunately, there are currently no H atom sticking experiments where mono-energetic H atoms have been sent onto a metal surface to which these calculations could be compared. However, given the distinct differences between ODF and LDFA in terms of sticking verification of one of those models should be feasible through future experiments.

\bibliography{mylib2}